\documentclass[twocolumn,showpacs,preprintnumbers,amsmath,amssymb,nofootinbib]{revtex4-1}
\usepackage{hyperref}
\usepackage{graphicx}
\usepackage{tikz}

\begin{document}

\title{On the effect of the cosmological expansion on the gravitational lensing by a point mass}
\author{Oliver F. Piattella}
\email{oliver.piattella@pq.cnpq.br}
\affiliation{Physics Department, Universidade Federal do Esp\'irito Santo, Vit\'oria, 29075-910, Brazil}

\date{\today}

\begin{abstract}
 We analyze the effect of the cosmological expansion on the deflection of light caused by a point mass, adopting the McVittie metric as the geometrical description of a pointlike lens embedded in an expanding universe. In the case of a generic, non-constant Hubble parameter $H$ we derive and approximately solve the null geodesic equations, finding an expression for the bending angle $\delta$, which we expand in powers of the mass-to-closest approach distance ratio and of the impact parameter-to-lens distance ratio. It turns out that the leading order of the aforementioned expansion is the same as the one calculated for the Schawarzschild metric and that cosmological corrections contribute to $\delta$ only at sub-dominant orders. We explicitly calculate these cosmological corrections for the case of $H$ constant and find that they provide a correction of order $10^{-11}$ on the lens mass estimate.
 \end{abstract}


\maketitle


\section{Introduction}

The effect of the cosmological constant $\Lambda$ (and thus, by extension, of cosmology) on the bending of light is an issue which has raised interest since a pioneering work by Rindler and Ishak in 2007 \cite{Rindler:2007zz}. The common intuition is that $\Lambda$ cannot have any local effect on the deflection of light because it is homogeneously distributed in the universe, thus not forming lumps which may act as lenses. Moreover, assume General Relativity (GR) as the fundamental theory of gravity, and consider Kottler metric \cite{kottler1918physikalischen}:\footnote{We use throughout this paper $G = c = 1$ units.}
\begin{equation}
	ds^2 = f(r)dt^2 - f^{-1}(r)dr^2 - r^2d\Omega^2\;,
\end{equation}
where $d\Omega^2 = d\theta^2 + \sin^2\theta d\phi^2$ and:
\begin{equation}\label{fKottler}
	f(r) \equiv 1 - \frac{2M}{r} - \frac{\Lambda r^2}{3}\;,
\end{equation}
as the description of a point mass $M$ embedded in a de Sitter space. It turns out that $\Lambda$ does not appear in the null geodesic trajectory equation, cf. e.g. Eq.~(17) of  Ref.~\cite{islam1983cosmological}. Indeed:
\begin{equation}\label{nullgeodequ}
	\frac{d^2u}{d\phi^2} + u = 3Mu^2\;, 
\end{equation}
where $u = 1/r$ is the inverse of the radial distance from the lens. Therefore, one may conclude that $\Lambda$ does not affect the bending of light, which is then entirely due to the presence of the point mass $M$.

On the other hand, in Ref.~\cite{Rindler:2007zz} the authors point out that the bending angle cannot be calculated as the angle between the asymptotic directions of the light ray, since these do not exist. Indeed, from Eq.~\eqref{fKottler} one sees that $r \le \sqrt{3/\Lambda}$, i.e. a cosmological horizon exists. In other words, the effect of $\Lambda$ enters in the boundary conditions that we choose when solving Eq.~\eqref{nullgeodequ}.
 
Thus, the authors of Ref.~\cite{Rindler:2007zz} find that $\Lambda$ enters the definition of the deflection angle in the following way [cf. their Eq. (17)]:
\begin{equation}\label{RindlerIshakResult}
	\psi_0 \approx \frac{2M}{R}\left(1 - \frac{2M^2}{R^2} - \frac{\Lambda R^4}{24M^2}\right)\;,
\end{equation}
where $R$ is the closest approach distance. Twice $\psi_0$ is the deflection angle. Therefore, one identifies the well-known Schwarzschild contribution $4M/R$, weighed by $\Lambda$, which tends to thwart the deflection.

After Ref.~\cite{Rindler:2007zz}, many authors confirmed with their calculations that $\Lambda$ does enter the formula for the deflection angle, although sometimes in a way different from the one in Eq.~\eqref{RindlerIshakResult}. See e.g. Refs.~\cite{Schucker:2007ut, Ishak:2008ex, Ishak:2008zc, Sereno:2008kk, Ishak:2010zh, Biressa:2011vy, Hammad:2013wda, Arakida:2016bvr}.

On the other hand, there are few works which do not agree with the above mentioned results, see e.g. \cite{Park:2008ih, Khriplovich:2008ij, Simpson:2008jf, Butcher:2016yrs}. The main criticism is that the Hubble flux is not properly taken into account, i.e. the relative motion among source, lens and observer is neglected. In particular, the authors of Ref.~\cite{Simpson:2008jf} argue that the $\Lambda$ contribution in Eq.~\eqref{RindlerIshakResult} is cancelled by the aberration effect due to the cosmological relative motion. Another interesting remark made in Ref.~\cite{Simpson:2008jf} is that the contribution of $\Lambda$ to the deflection angle does not vanish for $M \to 0$ in Eq.~\eqref{RindlerIshakResult}. In this respect, consider also e.g. Eq.~(25) of Ref.~\cite{Biressa:2011vy}:
\begin{eqnarray}\label{Pachecodelta}
	\delta &=& \frac{4M}{b} - Mb\left(\frac{1}{r_S^2} + \frac{1}{r_O^2}\right) + \frac{2Mb\Lambda}{3} - \frac{b\Lambda}{6}(r_S + r_O)\nonumber\\ &-& \frac{b^3\Lambda}{12}\left(\frac{1}{r_S} + \frac{1}{r_O}\right) + \frac{Mb^3\Lambda}{6}\left(\frac{1}{r_S^2} + \frac{1}{r_O^2}\right) + \cdots
\end{eqnarray}
Here $b$ is the impact parameter and $r_S$ and $r_O$ are the radial distances from the lens to the source and to the observer, respectively. Taking the limit $M \to 0$ in the above equation does not imply $\delta \to 0$. 

This seems to be odd since we do not expect lensing without a lens. However, this is the result that one obtains when the Hubble flux is not taken into account. Indeed, the author of Ref.~\cite{Butcher:2016yrs} constructs ``by hand" cosmological observers in the Kottler metric and finds that $\Lambda$ has no observable effect on the deflection of light. 

Therefore, according to the results of Refs.~\cite{Park:2008ih, Khriplovich:2008ij, Simpson:2008jf, Butcher:2016yrs}, the standard approach to gravitational lensing does not need modifications. For the sake of clarity, the standard approach to gravitational lensing consists in using the result on the deflection angle obtained from the Schwarzschild metric (which models the lens) together with the cosmological angular diameter distances calculated from the Friedmann-Lema\^itre-Robertson-Walker (FLRW) metric. See e.g. Ref.~\cite{Weinberg:2008zzc}.

In Ref.~\cite{Piattella:2015xga} we also tackled the investigation of whether a cosmological constant might affect the gravitational lensing by adopting the McVittie metric \cite{McVittie:1933zz} as the description of the lens. The McVittie metric is an exact spherical symmetric solution of Einstein equations in presence of a point mass and a cosmic perfect fluid. See e.g. Refs.~\cite{Nolan:1998xs, Nolan:1999kk, Nolan:1999wf, Kaloper:2010ec, Lake:2011ni, Nandra:2011ug, Nandra:2011ui, Nolan:2014maa, Faraoni:2015ula} for mathematical investigations of the geometrical properties of McVittie metric. In Ref.~\cite{Piattella:2015xga} we considered a constant Hubble factor, thus the geometry involved is the very Kottler one considered by most of the authors cited in this paper, but written in a different reference frame. Our results corroborate those of Refs.~\cite{Park:2008ih, Khriplovich:2008ij, Simpson:2008jf, Butcher:2016yrs}.

In the present paper we generalize the results of Ref.~\cite{Piattella:2015xga} to the case of a generic time-dependent $H$. See also Ref.~\cite{Aghili:2014aga}. This is necessary in order to make contact with the current standard model of cosmology, the $\Lambda$CDM model, in which pressureless matter prevents $H$ to be a constant. In particular, Friedmann equation for the $\Lambda$CDM model reads:
\begin{equation}
	\frac{H^2}{H_0^2} = \Omega_{\rm m}a^{-3} + \Omega_\Lambda\;,
\end{equation}
where $H_0$ is the Hubble constant, $a$ is the scale factor, $\Omega_{\rm m}$ is the present density parameter of pressureless matter and $\Omega_\Lambda = 1 - \Omega_{\rm m}$ is the density parameter of the cosmological constant. The time-derivative of $H$ can be easily computed as:
\begin{equation}
	\frac{\dot{H}}{H_0^2} = -\frac{3}{2}\Omega_{\rm m}a^{-3}\;.
\end{equation}
Since $\Omega_{\rm m} \approx 0.3$, one can see that $\dot{H}_0$ and $H_0^2$ are of the same order at present time. Moreover, $|\dot H| \ge H^2$ for $1 + z \ge \sqrt[3]{2\Omega_\Lambda/\Omega_{\rm m}}$. This gives a redshift $z > 1.67$. Many of the observed sources and lenses have redshifts larger than this limit,\footnote{See e.g. the CASTLES survey, \url{https://www.cfa.harvard.edu/castles/}.} therefore the above calculation shows that assuming $H$ constant is a very bad approximation and if one plans to make contact with observation it is necessary to go beyond the static case of the Kottler metric. The McVittie metric with a generic, non-constant $H$ provides an opportunity to do this.

Very recently, the deflection of light in a cosmological context with a generic $H(t)$ has been considered in Ref.~\cite{Faraoni:2016wae}. The main result is the following, cf. Eq.~(11) of Ref.~\cite{Faraoni:2016wae}:
\begin{equation}\label{Faraonieq}
	\Delta\tilde\varphi = \frac{4\tilde M_{\rm MSH}}{\tilde R} - 2H^2\tilde R^2\;. 
\end{equation}
Here $\tilde M_{\rm MSH}$ is the Misner-Sharp-Hernandez mass \cite{Misner:1964je, Hernandez:1966zia} contained in a radius $\tilde R = a(t)r$. Hence, it turns out that the effect of cosmology on the gravitational lensing depends of whether one takes into account the total contribution (local plus cosmological) to the mass or just the local one. However, in both cases $\tilde M_{\rm MSH}$ can be decomposed in the local $m$ contribution plus the cosmological one, which is cancelled by the $- 2H^2\tilde R^2$ in the above Eq.~\eqref{Faraonieq}. So, it appears that the net result is that the cosmic fluid does not contribute directly to the gravitational lensing.

As a final remark, we must stress that the McVittie metric is an extremely oversimplified model of an actual lens, which has a more complicated structure than that of a point. However, the results of our investigation may shed an important light and give valuable insight for a future research which takes into account a more complex structure of the lens. Up to our knowledge, lenses with structure different from a point have been considered only by Ref.~\cite{Biressa:2011vy}.

The present paper is structured as follows. In Sec.~\ref{Sec:McVittiemetric} we present the McVittie metric and its principal features. In Sec.~\ref{Sec:BendMcVittie} we obtain the null geodesic equations and calculate the deflection angle. In Sec.~\ref{Sec:lambdadominatedsubdomterm} we focus on the case of a constant $H$ and calculate exactly the subdominant contribution to the deflection angle, estimating a relative correction on the mass determination of about $10^{-11}$, due to the Hubble flux. In Sec.~V we present our conclusions. Throughout the paper we use $G = c = 1$ units.


\section{The McVittie metric}\label{Sec:McVittiemetric}

The McVittie metric \cite{McVittie:1933zz} can be written in the following form:
\begin{equation}\label{mcvittiecomoving}
	ds^2 = -\left(\frac{1 - \mu}{1 + \mu}\right)^2dt^2 + (1 + \mu)^4a(t)^2(d\rho^2 + \rho^2d\Omega^2)\;,
\end{equation}
where $a(t)$ is the scale factor, $d\Omega^2 = d\theta^2 + \sin^2\theta d\phi^2$ and
\begin{equation}\label{mudefinition}
	\mu \equiv \frac{M}{2a(t)\rho}\;,
\end{equation}
where $M$ is the mass of the point. One can check that for $a =$ constant the Schwarzschild metric in isotropic coordinates is recovered, whereas for $M = 0$ the FLRW metric is recovered.
 
When $\mu \ll 1$, the McVittie metric~\eqref{mcvittiecomoving} can be approximated by:
\begin{equation}\label{mcvittiepert}
	ds^2 = -\left(1 - 4\mu\right)dt^2 + (1 + 4\mu)a(t)^2(d\rho^2 + \rho^2d\Omega^2)\;,
\end{equation}
i.e. it takes the form of a perturbed FLRW metric in the Newtonian gauge with gravitational potential $2\mu$. Since there is only a single gravitational potential, then no anisotropic pressure is present \cite{Weinberg:2008zzc, Dodelson:2003ft}. 

Calculating the Einstein tensor from the McVittie line element \eqref{mcvittiecomoving}, one gets:
\begin{equation}
	G^t{}_t = 3H^2\;, \quad G^r{}_r = G^\theta{}_\theta = G^\phi{}_\phi = 3H^2 + \frac{2\dot{H}(1 + \mu)}{1 - \mu}\;,
\end{equation}
from which one deduces that the pressure of the cosmological medium has the following form:
\begin{equation}
	P = -\frac{1}{8\pi}\left[3H^2 + \frac{2\dot{H}(1 + \mu)}{1 - \mu}\right]\;,
\end{equation}
i.e. it is not homogeneous and diverging when $\mu = 1$. If $H =$ constant, then there is no divergence and the pressure is also a constant. This is the case of the Schwarzschild-de Sitter space, described by Kottler metric \cite{kottler1918physikalischen}. When $\dot H \neq 0$, faraway from the point mass, i.e. for $\mu \ll 1$, one gets the usual result of cosmology:
\begin{equation}
	P = -\frac{1}{8\pi}\left(3H^2 + 2\dot{H}\right) + \mathcal{O}(\mu) = -\rho - \frac{\dot{H}}{4\pi} + \mathcal{O}(\mu)\;,
\end{equation}
i.e. the acceleration equation. Isotropy is preserved since $G^r{}_r = G^\theta{}_\theta = G^\phi{}_\phi$, i.e. there is no anisotropic pressure, as we already mentioned.

Following Faraoni \cite{Faraoni:2015ula}, but also Park \cite{Park:2008ih}, McVittie metric~\eqref{mcvittiecomoving} can be reformulated in terms of the areal radius
\begin{equation}
	R = a\rho (1 + \mu)^2\;,
\end{equation}
and gets the following form:
\begin{eqnarray}\label{arealradiuscoord}
	ds^2 = -\left(1 - \frac{2M}{R} - H^2R^2\right)dt^2 + \frac{dR^2}{1 - \frac{2M}{R}}\nonumber\\ - \frac{2HR}{\sqrt{1 - \frac{2M}{R}}}dtdR + R^2d\Omega^2\;.
\end{eqnarray}
Changing the time coordinate as
\begin{equation}
	F(r,t)dT = dt + \frac{HR}{\sqrt{1 - \frac{2M}{R}}(1 - \frac{2M}{R} - H^2R^2)}dR\;,
\end{equation}
the above line element \eqref{arealradiuscoord} can be finally cast as
\begin{eqnarray}\label{propcoords}
	ds^2 = -\left(1 - \frac{2M}{R} - H^2R^2\right)F^2dT^2\nonumber\\ + \frac{dR^2}{1 - \frac{2M}{R} - H^2R^2} + R^2d\Omega^2\;.
\end{eqnarray}
If $H$ is constant, then $F$ can be set to unity and we recover Kottler metric. Using Eq.~\eqref{propcoords} and calculating the Misner-Sharp-Hernandez mass \cite{Misner:1964je, Hernandez:1966zia} of a sphere of proper radius $R$, one finds \cite{Faraoni:2015ula}:
\begin{equation}\label{MSHmassmcVittie}
	m_{\rm MSH} = M + \frac{H^2R^3}{2} = M + \frac{4\pi}{3}\rho R^3\;,
\end{equation} 
which contains the time-independent contribution $M$ from the point mass plus the mass of the cosmic fluid contained in the sphere. Therefore, $M$ has indeed the physical meaning of the mass of the point. In the Kottler case one can also define a Komar integral, or Komar mass, and verify that it is indeed equal to $M$ \cite{Kastor:2008xb}.


\section{The bending of light in the McVittie metric}\label{Sec:BendMcVittie}

We now revisit the calculation for the bending angle performed in Ref.~\cite{Piattella:2015xga}, but taking into account a general non-constant Hubble factor $H(t)$. We perform the calculations in two different ways: the one in this section is also used in Ref.~\cite{Piattella:2015xga} and is based on the approach usually adopted to study weak lensing, see e.g. Ref.~\cite{Dodelson:2003ft, Bartelmann:1999yn}. In this approach the origin of the coordinate system is occupied by the observer. The second way is the one in which the lens is put at the origin of the coordinate system and it is employed in appendix~\ref{App:usuallens}.

We adopt $\mu$ as perturbative parameter and work at the first order approximation in $\mu$. The observed angle of a lensed source is of the order of the arcsecond, which corresponds to $\theta_O \approx 10^{-6}$ radians. See e.g. the CASTLES survey lens database.\footnote{\url{https://www.cfa.harvard.edu/castles/}} At least for Einstein ring systems, the bending angle is of the same order of the observed angle, i.e. $\delta \approx 10^{-6}$. But, at the same time, the bending angle is of the same order of $\mu$. Therefore, we draw the conclusion that $\mu \approx 10^{-6}$ and the truncation error, when working at first order in $\mu$, is $\mathcal{O}(10^{-12})$.

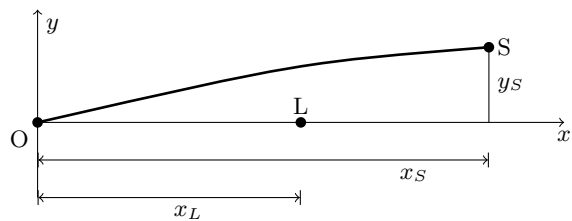
\begin{figure}[h!]
\centering
	\begin{tikzpicture}
	\draw[->] (0,0) -- (7,0) node[below] {$x$};
	\draw[->] (0,-1) -- (0,1.5) node[below right] {$y$};
	\draw (6,0) -- (6,1);
	\draw (6,0.5) node[right] {$y_S$};
	\fill (0,0) circle (2pt) node[below left] {O};
	\fill (3.5,0) circle (2pt) node[above] {L};
	\fill (6,1) circle (2pt) node[right] {S};
	\draw[line width=1pt] (0, 0) .. controls(3.5,0.8) .. (6, 1);
	\draw[|<->|] (0,-0.5) -- (6,-0.5);
	\draw[|<->|] (3.5,-1) -- (0,-1);
	\draw (5,-0.5) node[below] {$x_S$};
	\draw (2,-1) node[below] {$x_L$};
\end{tikzpicture}
\caption{Scheme of lensing.}
\label{figuretrajectory}
\end{figure}

The geometry of the lensing process is depicted in Fig.~\ref{figuretrajectory}. In this scheme, the observer stays at the origin of the spatial coordinate system and $x$ is the comoving coordinate along the observer-lens axis. 

The observer has spatial position $(0,0,0)$ and the lens has spatial position $(x_L,0,0)$. The McVittie metric \eqref{mcvittiecomoving} written in Cartesian spatial coordinates is the following:
\begin{equation}\label{mcvittiecomovingCart}
	ds^2 = -\left(\frac{1 - \mu}{1 + \mu}\right)^2dt^2 + (1 + \mu)^4a(t)^2\delta_{ij}dx^idx^j\;.
\end{equation}
Note that the spherical symmetry of the McVittie metric implies rotational symmetry about the observer-lens axis. Therefore, we set $z = 0$ without losing generality and the source has thus spatial position $(x_S,y_S,0)$.

Since in metric~\eqref{mcvittiecomoving} the lens lays at the origin of the coordinate system, we have to perform a translation along the $x$ axis in the Cartesian coordinates of metric~\eqref{mcvittiecomovingCart}, so that $\mu$ gets the following form:
\begin{equation}\label{mudefcart}
	\mu = \frac{M}{2a(t)\sqrt{(x - x_L)^2 + y^2 + z^2}}\;.
\end{equation}
Introducing an affine parameter $\lambda$ and the four-momentum $P^\mu = dx^\mu/d\lambda$, we can derive from metric~\eqref{mcvittiecomovingCart} the following relation:
\begin{equation}\label{P0}
	g_{\mu\nu}P^\mu P^\nu = 0 \quad \Rightarrow \quad P^0 = \frac{1 + \mu}{1 - \mu}p \sim (1 + 2\mu)p\;, 
\end{equation}
where $p^2 = g_{ij}P^iP^j$ is the proper momentum. The above equation represents the usual gravitational redshift experienced by a photon passing through the potential well generated by the point mass. Note that this potential well is not static since $\mu$ is time-dependent.

Now, we calculate the geodesic equations for the photon propagating in the McVittie metric~\eqref{mcvittiecomovingCart}:
\begin{equation}
	\frac{d^2x^\nu}{d\lambda^2} + \Gamma^\nu_{\alpha\beta}\frac{dx^\alpha}{d\lambda}\frac{dx^\beta}{d\lambda} = 0\;.
\end{equation}
The geodesic equation for $\nu = 0$ has the following form:
\begin{equation}\label{geodesicequation0}
	\frac{dP^0}{d\lambda} = 2\dot\mu(P^0)^2 + 4\mu_{,i}P^0P^i - p^2[H(1 + 4\mu) + 2\dot\mu]\;.
\end{equation}
Using Eq.~\eqref{P0} and the fact that, from Eq.~\eqref{mudefinition}, $\dot\mu = -H\mu$, one finds:
\begin{equation}
	p\frac{dp}{dt} = -Hp^2 + 2H\mu p^2 + 4\mu_{,i}P^ip\;.
\end{equation}
The zeroth-order term $Hp^2$ represents the usual cosmological redshift term. The spatial geodesics equations have the form:
\begin{equation}\label{spatialgeodesicseq}
	\frac{dP^i}{d\lambda} = \frac{4\delta^{il}\mu_{,l}p^2}{a^2} - 2HpP^i - 4P^iP^k\mu_{,k}\;. 
\end{equation}
We now look for an equation for the quantity $dy/dx \equiv \tan\theta$, which represents the slope of the line tangent to the photon trajectory. Note that $\theta$ is a physical angle because of the isotropic form of metric~\eqref{mcvittiecomovingCart}.

Since we can invert $x(\lambda)$ to $\lambda(x)$, being it a monotonic function, we can rewrite Eq.~\eqref{spatialgeodesicseq} for $y$ and change the variable to $x$:
\begin{eqnarray}
	P^x\frac{d}{dx}\left(P^x\frac{dy}{dx}\right) = 4\mu_{,y}[(P^x)^2 + (P^y)^2] \nonumber\\ - 2Ha(1 + 2\mu)\sqrt{(P^x)^2 + (P^y)^2}P^x\frac{dy}{dx}\nonumber\\ - 4\frac{dx}{dz}P^x[P^y\mu_{,y} + P^x\mu_{,x}]\;,
\end{eqnarray}
where we used the fact that
\begin{equation}
	p^2 = a^2(1 + 4\mu)[(P^x)^2 + (P^y)^2]\;.
\end{equation}	
Expanding the left hand side and using $P^y/P^x = dy/dx$, we obtain:
\begin{eqnarray}\label{d2yd2xeq}
	\frac{d^2y}{dx^2} +\frac{1}{P^x}\frac{dP^x}{dx}\frac{dy}{dx} = 4\mu_{,y}\left[1 + \left(\frac{dy}{dx}\right)^2\right]\nonumber\\ - 2Ha(1 + 2\mu)\sqrt{1 + \left(\frac{dy}{dx}\right)^2}\frac{dy}{dx}\nonumber\\ - 4\frac{dy}{dx}\left(\frac{dy}{dx}\mu_{,y} + \mu_{,z}\right)\;.
\end{eqnarray}
In order to determine the second term on the left hand side, we use Eq.~\eqref{spatialgeodesicseq} for $x$:
\begin{eqnarray}\label{spatialgeodesicseqx}
	\frac{1}{P^x}\frac{dP^x}{dx} = 4\mu_{,x}\left[1 + \left(\frac{dy}{dx}\right)^2\right]\\ - 2Ha(1 + 2\mu)\sqrt{1 + \left(\frac{dy}{dx}\right)^2} - 4\left(\frac{dy}{dx}\mu_{,y} + \mu_{,z}\right)\nonumber\;. 
\end{eqnarray}
Combining the two equations \eqref{d2yd2xeq} and \eqref{spatialgeodesicseqx}, we finally find:
\begin{equation}\label{trajectoryeq}
	\frac{d^2y}{dx^2} = 4\mu_{,y}\left[1 + \left(\frac{dy}{dx}\right)^2\right] - 4\mu_{,x}\left[1 + \left(\frac{dy}{dx}\right)^2\right]\frac{dy}{dx}\;.
\end{equation}
Let's discuss a little about the spatial derivation of $\mu$. From Eq.~\eqref{mudefcart}, we get:
\begin{equation}\label{dmudy}
	\mu_{,y} = -\mu\frac{y}{\sqrt{(x - x_L)^2 + y^2}}\;, 
\end{equation}
and
\begin{equation}
	\mu_{,x} = -\mu\frac{(x - x_L)}{\sqrt{(x - x_L)^2 + y^2}}\;. 
\end{equation}
Notice that, when $\mu = 0$, Eq.~\eqref{trajectoryeq} becomes:
\begin{equation}
	\frac{d^2y}{dx^2} = 0\;.
\end{equation}
i.e. the zeroth-order trajectory is, as expected, a straight line in comoving coordinates. 

We now make a second approximation: we assume $dy/dx = \tan\theta$ to be small. From the CASTLES survey we know that the observed angle $\theta_O$ is of the order of the arcsecond, which corresponds to $\theta_O \approx 10^{-6}$ radians. The latter is larger that the actual angular position of the source, say $\theta_S$, because of the lensing geometry, see e.g. Fig.~\ref{lensfig}. For this reason, we can assume $dy/dx$ to be small along all the trajectory. Since $dy/dx = \tan\theta$, then $dy/dx = \theta + \theta^3/3 + \cdots$. The truncation error is then of order $\mathcal{O}(\theta^3) \sim 10^{-18}$.


We consider Eq.~\eqref{trajectoryeq} up to the lowest order term, i.e.
\begin{equation}\label{trajectoryeqsimpl}
	\frac{d^2y}{dx^2} = 4\mu_{,y} + \mathcal{O}(\mu\theta)\;,
\end{equation}
where $\mathcal{O}(\mu\theta) \sim 10^{-12}$. Using Eq.~\eqref{dmudy} for the derivative $\mu_{,y}$, the above equation becomes:
\begin{equation}
	\frac{d^2y}{dx^2} = -\frac{2My}{a(x)\left[(x - x_L)^2 + y^2\right]^{3/2}}\;.
\end{equation}
Note that $a$ is a function of time, but inverting $x(t)$ we can write $a$ as a function of $x$. For simplicity, we normalize $x$, $y$ and $2M$ to $x_L$, thus obtaining:
\begin{equation}\label{fundeqy}
	\frac{d^2Y}{dX^2} = -\alpha\frac{Y}{a(X)\left[(X - 1)^2 + Y^2\right]^{3/2}}\;,
\end{equation}
where $Y \equiv y/x_L$, $X \equiv x/x_L$ and $\alpha \equiv 2M/x_L$. The above equation was already found in Ref.~\cite{Piattella:2015xga}. Since $dY/dX = \tan\theta$ and $\tan\theta \sim \theta$, we can cast the above equation in the following form
\begin{equation}\label{fundeqtheta}
	\frac{d\theta}{dX} = -\alpha\frac{Y}{a(X)\left[(X - 1)^2 + Y^2\right]^{3/2}}\;.
\end{equation}
We \textit{define} the bending angle as follows:
\begin{equation}\label{bendangleintegral}
	\delta \equiv \int_{\theta_S}^{\theta_O}d\theta = -\alpha\int_{X_S}^0\frac{Y(X)dX}{a(X)\left[(X - 1)^2 + Y(X)^2\right]^{3/2}}\;,
\end{equation}
i.e. as the variation of the slope of the trajectory between the source and the observer.

We shall solve the above equation keeping the first order in $\alpha$. The order of magnitude of $\alpha$ can be estimated as follows:
\begin{equation}\label{alphaestimate}
	\alpha \equiv \frac{2M}{x_L} \approx 2MH_0/z_L\;,
\end{equation}
where we assumed a small redshift $z_L$. The above approximation becomes an exact result in the case of a constant Hubble parameter, see e.g. \eqref{achiHconst}. 

Therefore, $\alpha$ is proportional to the ratio between the Schwarzschild radius of the lens and the Hubble radius. This is $H_0M \sim 10^{-12}$ for a galaxy of $10^{10}$ M$_\odot$ and $H_0M \sim 10^{-9}$ for a cluster of $10^3$ galaxies each of mass $10^{10}$ M$_\odot$.

We now devote a small paragraph to the zeroth-order solution.

\subsection{The zeroth-order solution}

\begin{figure}[h!]
\centering
	\begin{tikzpicture}
	\draw[->] (0,0) -- (7,0) node[below] {$x$};
	\draw[->] (0,-0.5) -- (0,1.5) node[below right] {$y$};
	\draw (6,0) -- (6,1);
	\draw (6,0.5) node[right] {$y_S$};
	\fill (0,0) circle (2pt) node[below left] {O};
	\fill (6,1) circle (2pt) node[above] {S};
	\draw[line width=1pt] (0, 0.5) -- (6, 1);
	\draw[|<->|] (0,-0.5) -- (6,-0.5);
	\draw (3.5,-0.5) node[below] {$x_S$};
	\draw (0,0.25) node[right] {$y_S - \theta_Sx_S$};
\end{tikzpicture}
\caption{Zeroth-order solution.}
\label{figure-zeroorder}
\end{figure}
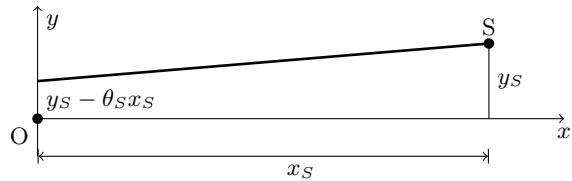

The zeroth-order solution (i.e. the one for $\alpha = 0$) of Eq.~\eqref{fundeqy} is a straight line in comoving coordinates:
\begin{equation}\label{zerothsol}
	y = \theta_S(x - x_S) + y_S\;,
\end{equation}
where $\theta_S \ll 1$ is the slope of the trajectory and $(x_S, y_S)$ are the comoving coordinates of the source. See Fig.~\ref{figure-zeroorder}.

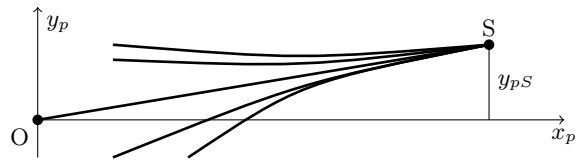
\begin{figure}[h!]
\centering
	\begin{tikzpicture}
	\draw[->] (0,0) -- (7,0) node[below] {$x_p$};
	\draw[->] (0,-0.5) -- (0,1.5) node[below right] {$y_p$};
	\draw (6,0) -- (6,1);
	\draw (6,0.5) node[right] {$y_{pS}$};
	\fill (0,0) circle (2pt) node[below left] {O};
	\fill (6,1) circle (2pt) node[above] {S};
	\draw[line width=1pt] (0, 0) -- (6, 1);
	\draw[line width=1pt] (1, 1) .. controls(3.5,0.8) .. (6, 1);
	\draw[line width=1pt] (1, 0.8) .. controls(3.5,0.7) .. (6, 1);
	\draw[line width=1pt] (1, -0.5) .. controls(3.5,0.5) .. (6, 1);
	\draw[line width=1pt] (2, -0.5) .. controls(3.5,0.5) .. (6, 1);
\end{tikzpicture}
\caption{Zeroth-order solution, using proper distances.}
\label{figure-zeroorder-proper}
\end{figure}

When we pass from comoving to proper distances by multiplying by the scale factor $a(x)$ we obtain:
\begin{equation}\label{phystraj}
	y_p = \theta_S x_p + \frac{a(x_p)}{a_S}(y_{pS} - \theta_Sx_{pS})\;,
\end{equation}
where we used a subscript $p$ to indicate the proper distance. The above is not a straight line trajectory, as also noticed by the authors of Ref.~\cite{Simpson:2008jf}. It is bent because of the $a(x_p)$ factor on the right hand side, whose effect vanishes only for $y_{pS} = \theta_Sx_{pS}$. The latter condition, when substituted in Eq.~\eqref{zerothsol}, represents the ray which gets to $y = 0$ when $x = 0$, i.e. the observed ray. See Fig.~\ref{figure-zeroorder-proper}.

The Hubble flux seems to bend away the trajectories such that we cannot detect any light. This happens isotropically, i.e. no observer could ever detect a bent ray but just the straight one coming directly from the source. 
 
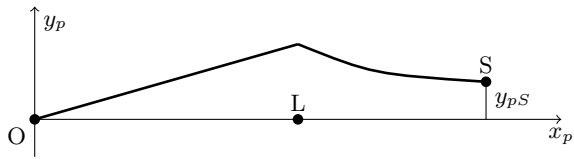
\begin{figure}[h!]
\centering
	\begin{tikzpicture}
	\draw[->] (0,0) -- (7,0) node[below] {$x_p$};
	\draw[->] (0,-0.5) -- (0,1.5) node[below right] {$y_p$};
	\draw (6,0) -- (6,0.5);
	\draw (6,0.25) node[right] {$y_{pS}$};
	\fill (0,0) circle (2pt) node[below left] {O};
	\fill (3.5,0) circle (2pt) node[above] {L};
	\fill (6,0.5) circle (2pt) node[above] {S};
	\draw[line width=1pt] (3.5, 1) .. controls(4.5,0.6) .. (6, 0.5);
	\draw[line width=1pt] (0, 0) -- (3.5, 1);
\end{tikzpicture}
\caption{A cosmologically bent ray, bent back. The ``back-bending".}
\label{figure-zeroorder-backbending}
\end{figure}

On the other hand, let's speculate about the following. If a cosmologically bent ray passes sufficiently close to a lens, then its trajectory could be bent back by the gravitational field of the lens, possibly allowing us to detect it. See Fig.~\ref{figure-zeroorder-backbending}.

This ``back-bending" seems to suggest that the bending angle must increase and therefore cosmology must somehow enter the gravitational lensing phenomenon.


\subsection{Calculation of the bending angle}

We now integrate Eq.~\eqref{bendangleintegral} retaining the first order only in $\alpha$. For this reason, the $Y(X)$ entering the integral is the zeroth-order solution, which we discussed in the previous subsection.

Since we are working at the first order in $\alpha$ we can assume without losing generality that the zeroth-order trajectory is horizontal, i.e. $Y^{(0)} = Y_S \equiv y_S/x_L$. 

The equation for the slope, i.e. Eq.~\eqref{fundeqtheta}, becomes
\begin{equation}\label{thetaeq}
	\frac{d\theta}{dX} = -\frac{\alpha Y_S}{a(X)\left[(X - 1)^2 + Y_S^2\right]^{3/2}}\;.
\end{equation}
In order to determine $a(X)$, we take advantage of metric~\eqref{mcvittiecomovingCart} and write:
\begin{equation}\label{axeq}
	dx^2 = \frac{1 - 8\mu}{1 + (dy/dx)^2}\frac{dt^2}{a^2}\;,
\end{equation}
which is a very complicated integration to perform, since it includes the very trajectory we want to determine. On the other hand, we are staying at the lowest possible order of approximation, therefore:
\begin{equation}
	dx = -\frac{dt}{a}\;,
\end{equation}
i.e. all the contributions coming from $\mu$ and $\theta$ of Eq.~\eqref{axeq} are of negligible order in Eq.~\eqref{thetaeq}. 

Now, write Eq.~\eqref{thetaeq} as follows:
\begin{equation}\label{thetaeq2}
	d\theta = -\frac{\alpha dX}{a(X)Y_S^2\left[1 + \frac{(X - 1)^2}{Y_S^2}\right]^{3/2}}\;.
\end{equation}
When $(X - 1)^2 \gg Y_S^2$ the above integration, whatever function $a$ might be of $X$, is $\mathcal{O}(Y_S)$. On the other hand, when $(X - 1)^2 \ll Y_S^2$ the above integration is $\mathcal{O}(1/Y_S^2)$.

Therefore, the main contribution comes from $X = 1$ and spans the interval $1 - Y_S < X < 1 + Y_S$. That is, most of the deflection takes place very close to the lens, as it happens for the case of the Schwarzschild metric. For this reason, we also approximate $a(X)$ with $a_L$, which is the scale factor when $x = x_L$.

Therefore, we end up with the following bending angle:
\begin{equation}\label{bendangle}
	\delta = -\frac{\alpha}{a_LY_S^2}(-2Y_S) = \frac{2\alpha}{a_LY_S} = \frac{4M}{a_Ly_S}\;.
\end{equation}
The above formula is general, valid for any kind of Hubble flow. We derive it using another method in Appendix~\ref{App:usuallens} and prove its validity in the case of a dust-dominated universe, for which an exact calculation is possible, in Appendix~\ref{App:bendingangledust}.

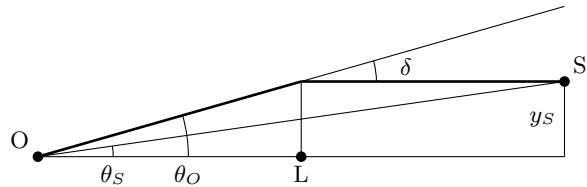
\begin{figure}[h!]
\centering
	\begin{tikzpicture}
	\draw (0,0) -- (7,0);
	\draw (7,0) -- (7,1);
	\draw [line width=1pt] (3.5,1) -- (7,1);
	\draw [line width=1pt] (0,0) -- (3.5,1);
	\draw (3.5,0) -- (3.5,1);
	\fill (0,0) circle (2pt) node[above left] {O};
	\fill (3.5,0) circle (2pt) node[below] {L};
	\fill (7,1) circle (2pt) node[above right] {S};
	\draw (0,0) -- (7, 2);
	\draw (0,0) -- (7, 1);
	\draw (4.5,1) arc (0:16:1);
	\draw (4.7,1.2) node[right] {$\delta$};
	\draw (1,0) arc (0:8:1);
	\draw (2,0) arc (0:16:2);
    \draw (1,0) node[below] {$\theta_S$};
    \draw (2,0) node[below] {$\theta_O$};
    \draw (7,0.5) node[left] {$y_S$};
\end{tikzpicture}
\caption{The thin-lens approximation.}
\label{lensfig}
\end{figure}

Now we apply formula~\eqref{bendangle} in the lens equation. Let us refer to Fig.~\ref{lensfig}. The geometry of this figure is justified by the fact that, as we showed earlier, the bending happens predominantly at the closest approach distance to the lens.

In Fig.~\ref{lensfig}, $\theta_S$ is the angular position of the source, so that $\theta_SD_{S}$ is the proper transversal position of the source, where $D_S$ is the angular-diameter distance from the observer to the source. The angle $\theta_O$ is the angular apparent position of the source, so that $\theta_O D_{S}$ is the transversal apparent position of the source.

Therefore, the lens equation in the thin-lens approximation can be written as:
\begin{equation}
	\theta_O D_S = \theta_S D_S + \delta D_{LS}\;,
\end{equation}
where $D_{LS}$ is the angular-diameter distance between lens and source. Using the result of Eq.~\eqref{bendangle}, we get
\begin{equation}
	\theta_O - \theta_S = \frac{4M}{a_Ly_S}\frac{D_{LS}}{D_S}\;.
\end{equation}
In the standard lens equation one has the closest approach distance to the lens, let's call it $R$, in place of $a_Ly_S$. One then writes $R = \theta_O D_L$ and thus finds the usual formula, see e.g. \cite{Weinberg:2008zzc}.

Now, since we found that the deflection occurs almost completely at the closest position to the lens, we can approximate $y_S \approx y_L$. Moreover, one also has $y_L = \theta_O x_L$ and $D_L = a_Lx_L$, from the definition of the angular-diameter distance to the lens. Thus, $a_Ly_S \approx a_L\theta_O x_L = \theta_O D_L$ and we recover the usual well-known formula:
\begin{equation}\label{usualresultlensing}
	\theta_O(\theta_O - \theta_S) = \frac{4M}{D_L}\frac{D_{LS}}{D_S}\;.
\end{equation}
Therefore, we can conclude that cosmology does not modify the bending angle at the leading order of the expansion in powers of $\mu$ and $\theta$. The cosmological ``drift" discussed earlier for the zeroth-order solution is already taken into account when using angular-diameter distances so that the final result does not change. 

However, sub-dominant terms do carry a cosmological signature, as we show in the next section. Here we address the simple case of a cosmological constant-dominated universe, where analytical calculations are possible.

\section{Next-to-leading order contributions to the bending angle in the case of a cosmological constant-dominated universe}\label{Sec:lambdadominatedsubdomterm}

As we saw in Eq.~\eqref{bendangle}, the leading contribution in the expansion for the bending angle calculated in the McVittie metric is the same as the one calculated for the Schwarzschild metric. Therefore, it is interesting to check if next-to-leading orders do carry a signature of the cosmological embedding of the point lens. We tackle this issue here in the case of a cosmological constant-dominated universe, for which exact calculations are possible, and leave a more general treatment as a future work.

When $H = H_0 = $ constant, one can find an analytic expression for $a(x)$:
\begin{equation}\label{achiHconst}
	x = \int_a^1\frac{da'}{H_0a^2} = \frac{1}{H_0}\left(\frac{1}{a} - 1\right) = \frac{z}{H_0}\;, 
\end{equation}
where in the last equality we introduced the redshift. The scale factor as function of the comoving distance is thus:
\begin{equation}
	\frac{1}{a(x)} = H_0x + 1 = H_0Xx_L + 1 = z_LX + 1\;,
\end{equation}
and Eq.~\eqref{thetaeq} becomes:
\begin{equation}\label{thetaeqHconst}
	\frac{d\theta}{dX} = -\frac{\alpha Y_S(z_LX + 1)}{\left[(X - 1)^2 + Y_S^2\right]^{3/2}}\;.
\end{equation}
As we anticipated, this equation can be solved exactly and the bending angle, as we defined it in Eq.~\eqref{bendangleintegral}, is the following:
\begin{eqnarray}
	\delta = \frac{\alpha}{Y_S}\left[\frac{1 + z_L + z_LY_S^2}{\sqrt{1 + Y_S^2}} +\right.\nonumber\\ \left. + \frac{(X_S - 1)(1 + z_L) - z_LY_S^2}{\sqrt{(X_S - 1)^2 + Y_S^2}}\right]\;.
\end{eqnarray} 
Expanding this solution for a small impact parameter $Y_S$ one gets:
\begin{eqnarray}
	\delta = \frac{2\alpha(1 + z_L)}{Y_S}\left[1 + \right. \nonumber\\ 
	\left.+ Y_S^2\frac{2(z_L - 1) + X_S[2 + X_S(z_L -1) - 4z_L]}{4(z_L + 1)(X_S - 1)^2}\right]\;, 
\end{eqnarray}
where we have already truncated $\mathcal{O}(Y_S^4)$ terms and put in evidence the leading order contribution $2\alpha(1 + z_L)/Y_S$, see Eq.~\eqref{bendangle}.

Recovering the physical quantities $Y_S = y_S/x_L$, $X_S = x_S/x_L$, $\alpha = 2M/x_L$ and using Eq.~\eqref{achiHconst} in order to express $x$ as the redshift, we get:
\begin{eqnarray}
	\delta = \frac{4M(1 + z_L)}{y_S}\left[1 + \right. \nonumber\\
	\left. \frac{y_S^2}{x_L^2}\frac{2z_L^2(z_L - 1) + z_S[2z_L + z_S(z_L -1) - 4z_L^2]}{4(z_L + 1)(z_S - z_L)^2}\right]\;. 
\end{eqnarray}
We already showed in the discussion leading to Eq.~\eqref{usualresultlensing} that $y_S \approx \theta_O x_L$, so that:
\begin{eqnarray}
	\delta = \frac{4M}{\theta_O D_L}\left[1 + \right. \nonumber\\
	\left. \theta_O^2\frac{2z_L^2(z_L - 1) + z_S[2z_L + z_S(z_L -1) - 4z_L^2]}{4(z_L + 1)(z_S - z_L)^2}\right]\;, 
\end{eqnarray}
and in the lens equation:
\begin{eqnarray}\label{lenseqsubdom}
	\theta_O(\theta_O - \theta_S) = \frac{4MD_{LS}}{D_LD_S}\left[1 + \right.\nonumber\\
	\left.\theta_O^2\frac{2z_L^2(z_L - 1) + z_S[2z_L + z_S(z_L -1) - 4z_L^2]}{4(z_L + 1)(z_S - z_L)^2}\right]\;.
\end{eqnarray}
Let's focus on Einstein ring systems, i.e. $\theta_S = 0$. We have in this case the mass estimate:\footnote{It is actually an estimate on the product $H_0M$, due to the presence of the angular-diameter distances. See Ref.~\cite{Weinberg:2008zzc}.}
\begin{eqnarray}\label{nexttoleadinordercorr}
	\frac{4MD_{LS}}{D_LD_S} = \theta_O^2\left[1 - \right. \nonumber\\ \left. \theta_O^2\frac{2z_L^2(z_L - 1) + z_S[2z_L + z_S(z_L -1) - 4z_L^2]}{4(z_L + 1)(z_S - z_L)^2}\right]\;.
\end{eqnarray}
The next-to-leading order correction is $\mathcal{O}(\theta_O^4)$ and depends on the redshifts of the lens and of the source.

Consider for example the Einstein ring Q0047-2808 of the CASTLES survey, for which $\theta_O = 2.7^{\prime\prime}$, $z_S = 3.60$ and $z_L = 0.48$. Substituting these numbers in Eq.~\eqref{nexttoleadinordercorr}, the correction on the mass estimate is therefore
\begin{equation}
	\frac{4MD_{LS}}{\theta_O^2D_LD_S} = 1 + 0.12\;\theta_O^2 = 1 + 2.03\cdot 10^{-11}\;.
\end{equation}
This is an extremely small correction which nonetheless depends on cosmology. Note that it is only one order of magnitude larger than the terms $\mathcal{O}(\mu^2)$ that we have neglected in our calculations.

\section{Conclusions}

We investigated whether cosmology affects the gravitational lensing caused by a point mass. To this purpose, we used McVittie metric as the description of the pointlike lens embedded in an expanding universe. The reason for this choice is to use a metric which properly takes into account the Hubble flux to which source, lens and observer are subject. We considered the general case in which the Hubble factor is a generic function of time and find that no contribution coming from cosmology enters the bending angle at the leading order, see Eq.~\eqref{bendangle}, thus strengthening the results obtained by \cite{Park:2008ih, Khriplovich:2008ij, Simpson:2008jf, Butcher:2016yrs}. 

We addressed the sub-dominant contributions to the bending angle in the special case of a constant Hubble factor $H = H_0$, for which exact calculations are possible. We found that in this case cosmology does affect the bending of light, through a combination of the lens and source redshifts, given in Eq.~\eqref{lenseqsubdom}. This correction is of order $10^{-11}$ for the Einstein ring Q0047-2808. 

We conclude that the standard approach to gravitational lensing on cosmological distances, which consists in patching together the results coming from Schwarzschild metric (which models the lens) and Friedmann-Lema\^itre-Robertson-Walker (FLRW) metric (which serves to calculate the cosmological angular diameter distances) does not require modifications.

Future developments of this investigation should address the entity subdominant orders of the expansion for the bending angle in a model-independent way. We expect the latter to depend on $H_L$, i.e. the Hubble parameter evaluated at the lens redshift. If these corrections were measurable, they might provide a new cosmological probe for determining the value of the Hubble parameter at different redshifts.

Another improvement would be that of tackling the analysis of the bending angle by constructing for the lens a density profile which could be more realistic than a Dirac delta (i.e. the one used here for a point mass). We expect that different lens density profiles would lead to different results in the mass estimates also from the point of view of the cosmological corrections, as showed in Ref.~\cite{Biressa:2011vy} for the case of the Kottler metric.


\acknowledgments{The author thanks CNPq (Brazil) for partial financial support. He is also indebted with D. Bacon, V. Marra and H. Velten for stimulating discussions and suggestions.}


\appendix

\section{Standard approach to gravitational lensing}\label{App:usuallens}

\begin{figure}[h!]
\centering
	\begin{tikzpicture}
	\draw[->] (0,0) -- (7,0);
	\draw[->] (2.5,-0.5) -- (2.5,1.5);
	\fill (0,0) circle (2pt) node[above left] {O};
	\fill (2.5,0) circle (2pt) node[below right] {L};
	\fill (6.5,0.5) circle (2pt) node[right] {S};
	\draw[line width=1pt] (0, 0) .. controls(2.5,1) .. (6.5, 0.5);
	\draw (1.5,0) node[below] {$\rho_L$};
	\draw (4,0.55) node[below] {$\rho_S$};
	\draw (2.5,0) -- (6.5,0.5);
	\draw (5,0) arc (0:18:1);
	\draw (5,0.17) node[right] {$\phi_S$};
\end{tikzpicture}
\caption{Scheme of lensing, with the lens at the origin of the coordinate system.}
\label{figuretrajectorypolarcoords}
\end{figure}
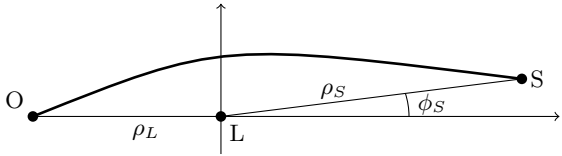

We now place the lens at the origin of the reference frame and use polar coordinates, as in Fig.~\ref{figuretrajectorypolarcoords}.

Again, we work at first order in $\mu$. For a photon, metric~\eqref{mcvittiepert} with $\theta = \pi/2$ gives:
\begin{equation}\label{McVittiephotonthetapi2}
	0 = -(1 - 4\mu)\dot{t}^2 + (1 + 4\mu)a^2\dot{\rho}^2 + (1 + 4\mu)a^2\rho^2\dot\phi^2\;,
\end{equation}
where the dot denotes derivation with respect to the affine parameter $\lambda$. Since $\xi_{(\phi)} = \delta^\mu_\phi\partial_\mu$ is a Killing vector, there exists the following conserved quantity:
\begin{equation}\label{Leq}
	L = g_{\mu\nu}P^\mu\xi^\nu_{(\phi)} = (1 + 4\mu)a^2\rho^2\dot\phi\;,
\end{equation}
where $P^\mu$ is the photon four-momentum. The geodesic equation for $t$, cf. Eq.~\eqref{geodesicequation0}, can be cast as follows:
\begin{equation}
	(1 - 4\mu)\ddot{t} - 4\frac{\partial\mu}{\partial\rho}\dot{t}\dot{\rho} + \frac{1}{a}\frac{da}{dt}(1 - 4\mu)\dot{t}^2 = 0\;,
\end{equation}
and written in the following compact form:
\begin{equation}\label{teqsimplified}
	\frac{d}{d\lambda}\left[(1 - 4\mu)a\dot{t}\;\right] + 4a\dot\mu\dot{t}^2 = 0\;.
\end{equation}
Recalling the definition of $\mu$ in Eq.~\eqref{mudefinition}, i.e. $\mu = M/(2a\rho)$, one can easily determine that $\dot\mu = -H\mu$. We neglect this contribution since indeed the ratio between the gravitational radius of the lens and the Hubble radius must be very small, as we discussed after Eq.~\eqref{alphaestimate}. 

Therefore, neglecting $HM$, Eq.~\eqref{teqsimplified} can be exactly integrated, giving the following result:
\begin{equation}\label{tdotsolution}
	\left(1 - \frac{2M}{a\rho}\right)\dot{t} = \frac{E}{a} + \mathcal{O}(HM)\;,
\end{equation}
where $E$ is an integration constant. We found a mixture of the known results for the Schwarzschild metric and for the FLRW one. Indeed, if $H = 0$ then $a$ is an unimportant constant which we can incorporate in the definitions of $\rho$ and $E$ and we recover the result for the Schwarzschild metric. On the other hand, with $M = 0$ we recover the usual cosmological decay of the energy of a photon, which is inversely proportional to the scale factor. 

Combining Eq.~\eqref{McVittiephotonthetapi2} with Eq.~\eqref{tdotsolution}, we can write the following equation for $\rho$:
\begin{equation}\label{dotrhoeq}
	\frac{a^4}{E^2}\dot\rho^2 = 1 - \frac{D^2}{\rho^2}(1 - 8\mu)\;,
\end{equation}
where $D \equiv L/E$ is a parameter associated to the closest approach distance $\rho_L$, defined as the one for which $\dot\rho_L = 0$, i.e.
\begin{equation}\label{closestapproach}
	\rho_L = D(1 - 4\mu_L)\;,
\end{equation}
where $\mu_L$ is $\mu$ evaluated at the closest approach distance, i.e. $\mu_L = M/(2a_L\rho_L)$.

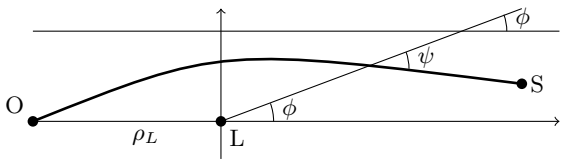
\begin{figure}[h!]
\centering
	\begin{tikzpicture}
	\draw[->] (0,0) -- (7,0);
	\draw[->] (2.5,-0.5) -- (2.5,1.5);
	\fill (0,0) circle (2pt) node[above left] {O};
	\fill (2.5,0) circle (2pt) node[below right] {L};
	\fill (6.5,0.5) circle (2pt) node[right] {S};
	\draw[line width=1pt] (0, 0) .. controls(2.5,1) .. (6.5, 0.5);
	\draw (1.5,0) node[below] {$\rho_L$};
	\draw (2.5,0) -- (6.5,1.5);
	\draw (3.2,0) arc (0:20:0.7);
	\draw (3.2,0.17) node[right] {$\phi$};
    \draw (0,1.2) -- (7,1.2);
    \draw (6.3,1.2) arc (0:20:0.64);
	\draw (6.3,1.37) node[right] {$\phi$};
	\draw (5,0.7) arc (0:20:0.65);
	\draw (5,0.85) node[right] {$\psi$};
\end{tikzpicture}
\caption{Schematic definition of the angle $\psi$, defined in Eq.~\eqref{psidef}. See also Fig.~2 of Ref.~\cite{Rindler:2007zz}.}
\label{fig:Rindlerbendangle}
\end{figure}

We use now the definition of the bending angle proposed by Rindler and Ishak in Ref.~\cite{Rindler:2007zz}, based on the following formula:
\begin{equation}\label{psidef}
	\tan\psi = \frac{\sqrt{g_{\phi\phi}}}{\sqrt{g_{\rho\rho}}}\left|\frac{d\phi}{d\rho}\right|\;,
\end{equation}
which represents the angle between the radial and the tangential directions of the photon trajectory, see Fig.~\ref{fig:Rindlerbendangle}. Using eqs.~\eqref{Leq} and \eqref{dotrhoeq} we find:
\begin{equation}
	\tan\psi = \frac{D}{\rho}\frac{1 - 4\mu}{\sqrt{1 - \frac{D^2}{\rho^2}(1 - 8\mu)}}\;.
\end{equation}
This expression can be rewritten in terms of the closest approach radius $\rho_L$ as follows:
\begin{eqnarray}\label{tanpsiformula}
	\tan\psi = \frac{\rho_L/\rho}{\sqrt{1 - \rho_L^2/\rho^2}}\left(1 - \frac{2M}{a\rho} + \frac{2M}{a_L\rho_L}\right)\;.
\end{eqnarray}
For $M = 0$ we obtain from Eq.~\eqref{tanpsiformula} that
\begin{equation}
	\tan\psi = \frac{\rho_L/\rho}{\sqrt{1 - \rho_L^2/\rho^2}} = \tan\phi\;,
\end{equation}
i.e. we recover the straight trajectory. Therefore, at any given position along the trajectory $\psi - \phi$ gives the local bending angle, i.e. the deviation from the straight-line trajectory. See Fig.~\ref{fig:Rindlerbendangle}.
The total bending angle is given by:
\begin{equation}
	\delta = \psi_S + \psi_O - \phi_S - \phi_O\;.
\end{equation}
If we assume $\rho_S$ and $\rho_O$ much larger than $\rho_L$, then the contributions from $\psi_S$ and $\psi_O$ are very small and practically negligible. Therefore, the dominant contribution to $\delta$ comes from $\phi_S + \phi_O$. In order to determine this sum, we must analyze the equation for the trajectory, i.e.
\begin{equation}
	\frac{d\rho}{d\phi} = \pm\rho\left(1 + \frac{4\mu - 4\mu_L}{1 - \rho_L^2/\rho^2}\right)\sqrt{\frac{\rho^2}{\rho_L^2} - 1}\;.
\end{equation}
For $\rho \gg \rho_L$, one can simplify this equation as follows: 
\begin{equation}
	\frac{d\rho}{d\phi} = \pm\frac{\rho^2}{\rho_L}\left(1 - 4\mu_L\right)\;,
\end{equation}
where we have considered only the leading-order correction to the equation for the straight line. The above equation tells us that the trajectory still is a straight line, far away from the lens, but tilted of an angle $4\mu_L$ from each side with respect to the horizontal. Therefore, the bending angle is
\begin{equation}
	\delta = 8\mu_L = \frac{4M}{a_L\rho_L}\;,
\end{equation}
which is identical to the result of Eq.~\eqref{bendangle} and also valid for a time-dependent $H$.

\section{Bending angle in a matter-dominated universe}\label{App:bendingangledust}

We check here formula~\eqref{bendangle} in the case of a matter-dominated universe, described by the Friedmann equation $H^2 = H_0^2/a^3$. The scale factor as a function of the comoving distance $x$ can be calculated as follows:
\begin{equation}\label{achiHmatter}
	x = \int_a^1\frac{da'}{H(a')a^2} = \frac{2}{H_0}(1 - \sqrt{a})\;, 
\end{equation}
which implies $a(x) = (1 - H_0x/2)^2$. With this $a(x)$, Eq.~\eqref{thetaeq} can be solved exactly and the bending angle is the following:
\begin{equation}
	\delta = \frac{2\alpha}{Y_S}\frac{4}{(H_0x_L - 2)^2} + \mathcal{O}(Y_S) = \frac{2\alpha}{a_LY_S} + \mathcal{O}(Y_S)\;,
\end{equation}
i.e. the same result found in Eq.~\eqref{bendangle}.


\bibliographystyle{unsrt}
\bibliography{McVittie}

\begin{thebibliography}{10}

\bibitem{Rindler:2007zz}
Wolfgang Rindler and Mustapha Ishak.
\newblock {Contribution of the cosmological constant to the relativistic
  bending of light revisited}.
\newblock {\em Phys. Rev.}, D76:043006, 2007.

\bibitem{kottler1918physikalischen}
Friedrich Kottler.
\newblock {\"U}ber die physikalischen grundlagen der einsteinschen
  gravitationstheorie.
\newblock {\em Annalen der Physik}, 361(14):401--462, 1918.

\bibitem{islam1983cosmological}
JN~Islam.
\newblock The cosmological constant and classical tests of general relativity.
\newblock {\em Physics Letters A}, 97(6):239--241, 1983.

\bibitem{Schucker:2007ut}
Thomas Schucker.
\newblock {Cosmological constant and lensing}.
\newblock {\em Gen.Rel.Grav.}, 41:67--75, 2009.

\bibitem{Ishak:2008ex}
Mustapha Ishak.
\newblock {Light Deflection, Lensing, and Time Delays from Gravitational
  Potentials and Fermat's Principle in the Presence of a Cosmological
  Constant}.
\newblock {\em Phys. Rev.}, D78:103006, 2008.

\bibitem{Ishak:2008zc}
Mustapha Ishak, Wolfgang Rindler, and Jason Dossett.
\newblock {More on Lensing by a Cosmological Constant}.
\newblock {\em Mon. Not. Roy. Astron. Soc.}, 403:2152--2156, 2010.

\bibitem{Sereno:2008kk}
M.~Sereno.
\newblock {The role of Lambda in the cosmological lens equation}.
\newblock {\em Phys. Rev. Lett.}, 102:021301, 2009.

\bibitem{Ishak:2010zh}
Mustapha Ishak and Wolfgang Rindler.
\newblock {The Relevance of the Cosmological Constant for Lensing}.
\newblock {\em Gen.Rel.Grav.}, 42:2247--2268, 2010.

\bibitem{Biressa:2011vy}
Tolu Biressa and J.A. de~Freitas~Pacheco.
\newblock {The Cosmological Constant and the Gravitational Light Bending}.
\newblock {\em Gen.Rel.Grav.}, 43:2649--2659, 2011.

\bibitem{Hammad:2013wda}
Fayçal Hammad.
\newblock {A note on the effect of the cosmological constant on the bending of
  light}.
\newblock {\em Mod. Phys. Lett.}, A28:1350181, 2013.

\bibitem{Arakida:2016bvr}
Hideyoshi Arakida.
\newblock {Effect of the Cosmological Constant on Light Deflection: Time
  Transfer Function Approach}.
\newblock {\em Universe}, 2(1):5, 2016.

\bibitem{Park:2008ih}
Minjoon Park.
\newblock {Rigorous Approach to the Gravitational Lensing}.
\newblock {\em Phys.Rev.}, D78:023014, 2008.

\bibitem{Khriplovich:2008ij}
I.~B. Khriplovich and A.~A. Pomeransky.
\newblock {Does Cosmological Term Influence Gravitational Lensing?}
\newblock {\em Int. J. Mod. Phys.}, D17:2255--2259, 2008.

\bibitem{Simpson:2008jf}
Fergus Simpson, John~A. Peacock, and Alan~F. Heavens.
\newblock {On lensing by a cosmological constant}.
\newblock {\em Mon. Not. Roy. Astron. Soc.}, 402:2009, 2010.

\bibitem{Butcher:2016yrs}
Luke~M. Butcher.
\newblock {Lambda does not Lens: Deflection of Light in the Schwarzschild-de
  Sitter Spacetime}.
\newblock 2016.

\bibitem{Weinberg:2008zzc}
Steven Weinberg.
\newblock {\em {Cosmology}}.
\newblock 2008.

\bibitem{Piattella:2015xga}
Oliver~F. Piattella.
\newblock {Lensing in the McVittie metric}.
\newblock {\em Phys. Rev.}, D93(2):024020, 2016.
\newblock [Erratum: Phys. Rev.D93,no.12,129901(2016)].

\bibitem{McVittie:1933zz}
G.C. McVittie.
\newblock {The mass-particle in an expanding universe}.
\newblock {\em Mon.Not.Roy.Astron.Soc.}, 93:325--339, 1933.

\bibitem{Nolan:1998xs}
Brien~C. Nolan.
\newblock {A Point mass in an isotropic universe: Existence, uniqueness and
  basic properties}.
\newblock {\em Phys. Rev.}, D58:064006, 1998.

\bibitem{Nolan:1999kk}
B.C. Nolan.
\newblock {A Point mass in an isotropic universe. 2. Global properties}.
\newblock {\em Class.Quant.Grav.}, 16:1227--1254, 1999.

\bibitem{Nolan:1999wf}
Brien~C. Nolan.
\newblock {A Point mass in an isotropic universe. 3. The region R less than or
  = to 2m}.
\newblock {\em Class. Quant. Grav.}, 16:3183--3191, 1999.

\bibitem{Kaloper:2010ec}
Nemanja Kaloper, Matthew Kleban, and Damien Martin.
\newblock {McVittie's Legacy: Black Holes in an Expanding Universe}.
\newblock {\em Phys.Rev.}, D81:104044, 2010.

\bibitem{Lake:2011ni}
Kayll Lake and Majd Abdelqader.
\newblock {More on McVittie's Legacy: A Schwarzschild - de Sitter black and
  white hole embedded in an asymptotically $\Lambda$CDM cosmology}.
\newblock {\em Phys. Rev.}, D84:044045, 2011.

\bibitem{Nandra:2011ug}
Roshina Nandra, Anthony~N. Lasenby, and Michael~P. Hobson.
\newblock {The effect of a massive object on an expanding universe}.
\newblock {\em Mon. Not. Roy. Astron. Soc.}, 422:2931--2944, 2012.

\bibitem{Nandra:2011ui}
Roshina Nandra, Anthony~N. Lasenby, and Michael~P. Hobson.
\newblock {The effect of an expanding universe on massive objects}.
\newblock {\em Mon. Not. Roy. Astron. Soc.}, 422:2945--2959, 2012.

\bibitem{Nolan:2014maa}
Brien~C. Nolan.
\newblock {Particle and photon orbits in McVittie spacetimes}.
\newblock {\em Class. Quant. Grav.}, 31(23):235008, 2014.

\bibitem{Faraoni:2015ula}
Valerio Faraoni.
\newblock {Cosmological and Black Hole Apparent Horizons}.
\newblock {\em Lect. Notes Phys.}, 907:pp.1--199, 2015.

\bibitem{Aghili:2014aga}
Mir~Emad Aghili, Brett Bolen, and Luca Bombelli.
\newblock {Effect of Accelerated Global Expansion on Bending of Light}.
\newblock 2014.

\bibitem{Faraoni:2016wae}
Valerio Faraoni and Marianne Lapierre-Leonard.
\newblock {Beyond lensing by the cosmological constant}.
\newblock 2016.

\bibitem{Misner:1964je}
Charles~W. Misner and David~H. Sharp.
\newblock {Relativistic equations for adiabatic, spherically symmetric
  gravitational collapse}.
\newblock {\em Phys. Rev.}, 136:B571--B576, 1964.

\bibitem{Hernandez:1966zia}
Walter~C. Hernandez and Charles~W. Misner.
\newblock {Observer Time as a Coordinate in Relativistic Spherical
  Hydrodynamics}.
\newblock {\em Astrophys. J.}, 143:452, 1966.

\bibitem{Dodelson:2003ft}
Scott Dodelson.
\newblock {\em {Modern cosmology}}.
\newblock 2003.

\bibitem{Kastor:2008xb}
David Kastor.
\newblock {Komar Integrals in Higher (and Lower) Derivative Gravity}.
\newblock {\em Class. Quant. Grav.}, 25:175007, 2008.

\bibitem{Bartelmann:1999yn}
Matthias Bartelmann and Peter Schneider.
\newblock {Weak gravitational lensing}.
\newblock {\em Phys. Rept.}, 340:291--472, 2001.

\end{thebibliography}

\end{document}